# Structure-Property Relation of $SrTiO_3$–$LaAlO_3$ Interfaces

Mark Huijben, Alexander Brinkman, Gertjan Koster, Guus Rijnders, Hans Hilgenkamp, and Dave H.A. Blank

*Faculty of Science and Technology and MESA[+] Institute for Nanotechnology, University of Twente, P.O. Box 217, 7500AE Enschede, The Netherlands*

A large variety of transport properties have been observed at the interface between the insulating oxides $SrTiO_3$ and $LaAlO_3$ such as insulation, 2D interface metallicity, 3D bulk metallicity, Kondo scattering, magnetism and superconductivity. The relation between the structure and the properties of the $SrTiO_3$-$LaAlO_3$ interface can be explained in a meaningful way by taking into account the relative 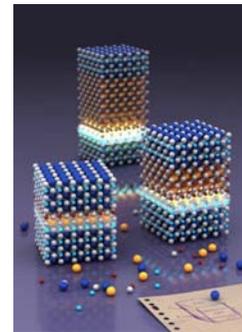 contribution of three structural aspects: oxygen vacancies, structural deformations (including cation disorder) and electronic interface reconstruction. The emerging phase diagram is much richer than for related bulk oxides due to the occurrence of interface electronic reconstruction. The observation of this interface phenomenon is a display of recent advances in thin film deposition and characterization techniques, and provides an extension to the range of exceptional electronic properties of complex oxides.

# 1. Introduction

Interface physics of strongly correlated oxides is a rapidly developing branch of materials science. In heterostructures of correlated oxide films, charge and spin states are reconstructed at the interfaces and hence affect the electronic and magnetic properties of the entire system. The ability to control both the atomic structure and composition of these oxide layers as well as their interfaces is emerging as one of the major challenges for the development of oxide-based electronic devices with a range of functional properties.

This recent trend in oxide research is a logical continuation of the seminal achievements in the exploration of novel properties of perovskite oxides. The large progress in the last decade was triggered by the possibility to produce well-defined single-terminated substrate surfaces [1,2] and to control the thin film growth, including pulsed laser deposition, on atomic scale using high-pressure reflective high energy electron diffraction. [3] These developments enabled the growth of epitaxial complex oxide heterostructures, such as multilayers and superlattices, with well defined interfaces. Furthermore, study of the structure and composition of complex oxides as well as their electronic structure is nowadays achievable by advanced analysis techniques, such as high-resolution surface x-ray diffraction techniques as well as scanning transmission electron microcopy in combination with electron energy loss spectroscopy. The recent developments of these highly advanced fabrication and analysis techniques have been crucial in investigating the growth-structure-property relationships of atomically controlled interfaces in oxides.

In 2004, Ohtomo and Hwang reported the first observation of a high-mobility electron gas at the interface between the two insulating perovskite oxides $LaAlO_3$ and $SrTiO_3$.[4] The fundamental mechanism underlying this new phenomenon at the oxide interface was proposed to be 'electronic reconstruction', where the spreading of charge across a polar/non-polar interface causes an effective electron doping at the interface. Quite soon, it became clear that the original picture had to be extended to include other effects, such as the formation of defects. Since then, an extensive amount of research has been conducted by various groups to unravel the nature of this tantalizing interface effect, both theoretically as well as experimentally.

Electronic reconstruction can intuitively be understood by considering the perovskite unit cell $ABO_3$ in terms of the constituting AO and $BO_2$ layers. For example, whereas $SrTiO_3$ and $LaAlO_3$ are seemingly similar, the $Sr^{2+}O^{2-}$ and $Ti^{4+}O^{2-}_2$ layers are charge-neutral, while in the ionic limit the charge states in the $LaAlO_3$ are positive for $La^{3+}O^{2-}$ and negative for $Al^{3+}O^{2-}_2$. In perovskite heterostructures the $AO-BO_2$ stacking sequence is maintained, and consequently a polarity discontinuity arises at the $LaAlO_3$-$SrTiO_3$ interface. To avoid a diverging potential build-up (with its associated large energy cost) charge has to be redistributed.[5] For conventional semiconductors, where the atoms have a fixed number of binding electrons, this rearrangement can be accomplished in the form of redistribution by compositional roughening.[6,7] For complex oxides, in some cases, the number of available binding electrons can vary (valence of the constituent ions can be multivalent) so that charge can be transferred across the interface at a lower energy cost than redistributing the ions. This results in the net transfer of electrons from $LaAlO_3$ to $SrTiO_3$ across the interface, see **Fig. 1a**. This intuitive picture provides an explanation for the observed electrical conduction. Theoretically, this concept has been refined by means of ab-initio calculations. Both calculations in the local density approximation [8-15] as well as by dynamical mean field theory [16,17] show that the electron density can be greatly enhanced at such oxide interfaces. Additionally, orbital reconstruction is predicted, as well as structural deformations and magnetic ordering. Whether the interface ground state is an antiferromagnetic insulator or metallic depends in a subtle way on the exact structure of the interface and the amount of defects.

In this progress report we present an overview of the experimental research carried out in the last few years on these conducting interfaces. The various experimental insights will be described with all the existing agreements and discrepancies. Section 2, in which the fabrication of the interfaces is described, will serve as a basis to understand the atomic ordering and relation between growth and structural properties. The large variety of observed transport properties at LaAlO$_3$/SrTiO$_3$ interfaces will be reviewed in Section 3, while we show in Section 4 how the transport properties are perceived to be related to the three main structural aspects of the interfaces: oxygen vacancies, structural deformations (including cation disorder), and electronic interface reconstruction. The emerging picture is one in which the relevance of each of these three aspects is determined by growth parameters, such as the partial oxygen pressure during deposition, and by tunable parameters such as applied electric field. This leads to a phase diagram in which all observed transport properties can be accommodated in a meaningful way and which surpasses related bulk oxides due to the interface electronic reconstruction. To our opinion this interface phenomenon is not only a scientific breakthrough,[18] but also provides an extension of the exceptional electronic properties of complex oxides to serve in novel concepts of oxide-electronic devices.

## 2. Fabrication of high-quality LaAlO$_3$/SrTiO$_3$ interfaces

### 2.1. Atomic interface ordering

The investigated materials, LaAlO$_3$ and SrTiO$_3$, are band insulators with bandgaps of ~5.6 eV and ~3.2 eV, respectively, and both belong to the perovskite structural family. The SrTiO$_3$ compound consists at room temperature of a simple cubic structure. The lattice parameters are 3.905 Å with the Ti atoms located at the corners and the Sr atoms at the centers of the cubes.[19,20] The oxygen atoms are placed at the centers of the twelve cube edges, giving corner-shared strings of TiO$_6$ octahedra, which extend in three dimensions. The TiO$_6$ octahedra are perfect with 90° angles and six equal Ti-O bonds at 1.952 Å. Each Sr atom is surrounded by twelve equidistant oxygen atoms at 2.760 Å. The SrTiO$_3$ compound undergoes a second-order phase transition from cubic (spacegroup $Pm\bar{3}m$) to tetragonal (spacegroup $I4/mcm$) at a temperature of ~110 K, due to rotation of neighbouring TiO$_6$ octahedra in opposite directions.[21-24] In the ionic limit strontium titanate can be described as Sr$^{2+}$Ti$^{4+}$O$^{2-}$$_3$.

On the other hand, the LaAlO$_3$ compound consists at room temperature of a rhombohedrally distorted perovskite structure (spacegroup $R\bar{3}c$), which undergoes a transition to the ideal cubic perovskite structure (spacegroup $Pm\bar{3}m$) at ~813 K.[25,26] The rhombohedral low temperature structure can be described as a perovskite structure with an antiphase rotation of the AlO$_6$ octahedra. This can be observed in diffraction analysis from a subtle splitting of the main peaks, attributed to a distortion from the cubic structure. However, this splitting is very small and can only be observed in high-resolution experiments. There have been many investigations of this phase transition [23,27-31] including recent ones using neutron powder methods to examine the thermal evolution of the structure.[32-34] The rhombohedral structure at room temperature can be described as a psuedocubic with lattice parameters of 3.791 Å with Al atoms located at the corners and the La atoms at the center of the cube.[34] This compound can be described in the ionic limit as La$^{3+}$Al$^{3+}$O$^{2-}$$_3$.

The charge states in the LaAlO$_3$ are positive for La$^{3+}$O$^{2-}$ and negative for Al$^{3+}$O$^{2-}$$_2$ On the contrary, for SrTiO$_3$, the Sr$^{2+}$O$^{2-}$ and Ti$^{4+}$O$^{2-}$$_2$ layers are charge-neutral. Since the perovskite heterostructures AO-BO$_2$ stacking sequence is maintained in heterostructures along the [001] direction, a polarity discontinuity arises at the LaAlO$_3$-SrTiO$_3$ interface. Since the Ti ion allows for mixed valence charge compensation, this results in the net transfer of electrons (nominally 0.5 electron per two-dimensional unit cell) from LaAlO$_3$ to SrTiO$_3$ across the interface, see Figure 1a. The extra electrons at the LaO-TiO$_2$ interface were confirmed by metallic conductivity and Hall measurements by Ohtomo and Hwang.[4,35] The interface charges at this 'n-type' interface are induced by electronic reconstruction conceivably through mixed-valence Ti states ($Ti^{4+}$ to $Ti^{3+}$) that place extra electrons in the $SrTiO_3$ conduction band. The analogous construction of the $AlO_2$-SrO interface, as shown in **Fig. 1b**, must now acquire extra holes per two-dimensional unit cell to maintain charge neutrality. This interface is formally called 'p-type'. Electrically, however, this interface was insulating.[4] As this p-charging is still conceivable and there are no available mixed valence states to accommodate the holes, an atomic reconstruction is required and will most likely be formed by the introduction of oxygen vacancies.

## 2.2. Substrate surface termination control

A prerequisite to obtain high-quality interfaces between $SrTiO_3$ and $LaAlO_3$ with control on the atomic scale is that the starting surface of the substrate has to be atomically smooth. However, for perovskites, the substrate surface obtained by cleaving or cutting, typically consists of an equal amount of AO- and $BO_2$-terminated domains separated by half unit-cell steps, see **Fig. 2a**. Thin film growth on these as-received substrates will result in an interface with a mixture of $LaO$-$TiO_2$ and $AlO_2$-SrO interfaces. To fabricate a single type interface, the initial substrate has to be single terminated by either AO or $BO_2$. When a polar material has charged layers, this becomes problematic. For example, surface studies on single crystalline $LaAlO_3$ substrates have given contradicting results about the surface termination by a thermal treatment.[36-39] Furthermore, surface reconstruction [40] is present for $La^{3+}Al^{3+}O_3$ crystals due to the presence of a polar surface of $(LaO)^+$ or $(AlO_2)^-$. Similar polar surfaces are always present for various other substrates, such as $Nd^{+3}Ga^{+3}O_3$, $K^{+1}Ta^{+5}O_3$ and $Dy^{+3}Sc^{+3}O_3$. As a result, only non-polar $SrTiO_3$ single crystalline substrates, expected to render single terminated surfaces with charge-neutral single surface terminations of either $TiO_2$ or SrO, are therefore used to investigate atomically controlled $LaAlO_3$/$SrTiO_3$ interfaces.

A chemical route was suggested to achieve this single termination for SrTiO$_3$ substrates by combining a chemical treatment and a thermal treatment.[1] The etching mechanism was later analyzed in more detail and a two-step chemical treatment was developed to form perfectly crystalline TiO$_2$-terminated SrTiO$_3$ surfaces,[2] see **Fig. 2b**. Recently a few refinements were reported with an additional etch procedure,[41] which was investigated by high-resolution synchrotron-radiation photoemission spectroscopy to result in very stable TiO$_2$-terminated surfaces.[42] Until now, no chemical treatments have been reported to produce the opposite single-terminated SrO surfaces, while heat treatment of the as-received SrTiO$_3$ substrates usually results in a mixed termination. The single-terminated SrO surfaces can, however, be obtained by deposition of a SrO monolayer on a single-terminated TiO$_2$ surface. Epitaxial growth of SrO has been reported to occur in a layer-by-layer mode for molecular beam epitaxy [43] as well as for pulsed laser deposition [44] at relatively low temperatures (400-500°C). For SrO monolayer growth, at normal SrTiO$_3$ deposition temperatures (850°C), pulsed laser *interval* deposition has to be applied.[45,46] In this deposition technique the total number of laser pulses for one monolayer has to be provided rapidly (50 Hz) to stabilize the correct SrO layer without multi-level islands[1]. This results in crystalline SrO-terminated SrTiO$_3$ surfaces with perfect straight step ledges, see **Fig. 2c**.

### 2.3. Growth of atomically controlled interfaces

Pulsed laser deposition has been used for the homoepitaxial growth of SrTiO$_3$ [1,2,47-51] as well as LaAlO$_3$. [52,53] High-quality heteroepitaxial growth of SrTiO$_3$ and LaAlO$_3$ has also been obtained by pulsed laser deposition, but very rarely by combining both materials before 2004. An example, where it was applied, is the growth of SrTiO$_3$ thin films on LaAlO$_3$ substrates to

---

[1] Concerning the deposition conditions, a single-crystal SrO target is ablated with an energy density of 1.3 J/cm$^2$. During growth, the substrate is held at 850 °C in an oxygen environment at 0.13 mbar.

produce electronically tunable microwave devices, such as resonators, filters and phase shifters. [54-59]

Following the initial publications of Ohtomo and Hwang, various groups have grown thin films of LaAlO$_3$ by pulsed laser deposition on single-terminated SrTiO$_3$ substrates to investigate the properties of the two possible heteroepitaxial interface configurations. To obtain well controllable layer-by-layer growth, it is suitable to utilize a single-crystal LaAlO$_3$ target. Most groups have used a KrF excimer laser at a repetition rate of 1 Hz and a laser fluency of 1 - 2 J/cm$^2$. A typical deposition temperature range is 750-850 $^\circ$C, while the oxygen pressure can be varied between $10^{-6}$ and $10^{-3}$ mbar to control the oxidation level, as will be discussed in more detail in section 4. The oxygen pressure has to be limited to this range to ensure the quality of the interface structure, because a transition from 2-dimensional layer-by-layer growth to island growth was observed for oxygen pressures of $10^{-2}$ mbar and higher. A very important role in the fabrication process is played by the annealing procedure after the thin film growth. To carefully study the oxidation level of the LaAlO$_3$-SrTiO$_3$ heterostructures, the oxygen pressure has to be kept at the deposition pressure during cool down. On the other hand, high oxygen pressure annealing has been used by various groups to fully oxidize the fabricated heterostructure and presumably to remove all oxygen vacancies. It must be noted that the oxygen pressure during growth determines the growth mode as well as the oxidation level. Subsequent exposure to high pressure molecular oxygen diminishes the number of oxygen vacancies, but a full stoichiometry is hard to achieve.

The surface quality was monitored by Reflection High-Energy Electron Diffraction (RHEED) during the growth of LaAlO$_3$ thin films on single TiO$_2$-terminated SrTiO$_3$ substrates as well as on single SrO-terminated SrTiO$_3$ substrates. The fluctuations in the RHEED intensity during the initial growth of the first unit cells are shown in **Fig. 3** for both types of surface terminations. Oscillations in the RHEED intensity can be observed in both cases, which indicate 2-dimensional layer-by-layer growth of LaAlO$_3$ for both types of SrTiO$_3$ surface terminations. The clear 2-dimensional spots in the RHEED pattern, shown in the insets, confirmed this growth behavior. The sharp decrease in RHEED intensity in both cases for the first LaAlO$_3$ unit cells can be explained by the difference in the optimal diffraction conditions for both materials, because the RHEED monitoring was initially aligned with the SrTiO$_3$ unit cell of the substrate. The difference in c-axis length between the initial SrTiO$_3$ unit cell (~3.905 Å) and the deposited LaAlO$_3$ unit cell (~3.791 Å) requires a new alignment of the RHEED monitoring for optimal surface analysis.

For well-aligned RHEED analysis, the 2-dimensional layer-by-layer growth of individual LaAlO$_3$ unit cells can be observed up to thicknesses of ~20 nm. The oscillations in the RHEED intensity were investigated to indicate growth of individual unit cells. The constant number of laser pulses, which is required to form one unit cell, and the constant RHEED intensity at the maximum of the oscillation suggest the growth of individual unit cells of LaAlO$_3$ with a constant surface roughness, see **Fig. 4a**. This was confirmed by the fluctuations in the full width at half maximum (FWHM) of the specular RHEED spot, which exhibit identical oscillations, but inverted compared to the specular spot amplitude. The constant FWHM value, after growth of each LaAlO$_3$ unit cell, indicates constant surface roughness without the formation of islands. The low level of surface roughness was confirmed by atomic force microscopy of a 26 unit cells (~10 nm) thick LaAlO$_3$ film on a TiO$_2$-terminated SrTiO$_3$ substrate, see **Fig. 4b**. The micrograph and the roughness analysis show smooth terraces with clear unit cell steps.

The quality of the LaAlO$_3$ thin films and the epitaxial relation to the SrTiO$_3$ substrates were investigated by x-ray diffraction for both types of heteroepitaxial interfaces (LaO-TiO$_2$ and AlO$_2$-SrO). In **Fig. 5** are θ-2θ scans shown for 26 unit cells thick LaAlO$_3$ thin films grown on SrTiO$_3$ substrates with a TiO$_2$-terminated surface as well as a SrO-terminated surface. In both cases only the (00*l*) reflections of the LaAlO$_3$ unit cell are present along with the reflections from the SrTiO$_3$ substrate, indicating c-axis growth. The LaAlO$_3$ unit cell is grown cube-on-cube on the SrTiO$_3$ unit cell with their a- and b-axes perfectly aligned. This results in an in-plane tensile strain in the thin film with in-plane lattice parameters similar to the substrate (3.905 Å). Consequently the c-axis lattice parameter of the LaAlO$_3$ thin film is shortened as compared to the bulk value (3.791 Å) and is ~3.73 Å for both types of heteroepitaxial interfaces. The presence of Kiessig fringes indicates a highly ordered crystalline structure between two well-defined smooth interfaces.

## 3. Transport properties

### 3.1. High-mobility electron gas

The structural properties of LaAlO$_3$ thin films grown on both types of single-terminated SrTiO$_3$ substrates are very similar, as was observed by RHEED and XRD. However, very large differences are present in their electronic properties. A convenient method to realize good ohmic contacts to the buried conducting layer at the interface is to provide wire-bonds that penetrate through the LaAlO$_3$ thin films. However, to enable careful analysis of the intrinsic interface transport properties, the samples have to be shielded from any light during the experiments and the 24 hours before to suppress the effects of possible photocarrier injection. The measured temperature dependence of the resistance is shown in **Fig. 6** for both types of heteroepitaxial interfaces. The difference in resistance at room temperature between both interfaces is a factor of ~10$^3$ and while the LaAlO$_3$ film on a TiO$_2$-terminated surface shows metallic behavior down to low temperatures, the LaAlO$_3$ film on a SrO-terminated surface shows insulating behavior and cannot be accurately measured at low temperatures. This distinct behavior between the two interfaces was already presented by Ohtomo and Hwang.[4] They noted that when a fraction of a monolayer of SrO was deposited on to the TiO$_2$-terminated SrTiO$_3$ substrate before LaAlO$_3$ growth, the carrier density decreased proportionally with increasing SrO coverage from 0 to 1 monolayer. This effect was investigated in more detail by Nishimura *et al.*, [60] where they varied the interfacial layer configuration between the two extreme cases, as given above, by inserting various fractional layers of SrO through a sliding mask technique to integrate several samples in a single experimental run. By changing the SrO coverage from 0 to 1, the electron density was controlled from a value corresponding to 0.5 electrons per Ti site to zero, see **Fig. 7**. This decrease in carrier density with SrO coverage came along with a systematic increase in sheet resistance. However, no systematic change could be observed in the temperature dependence of the mobility. All samples showed similar scattering behavior and the conductivity was just determined by the carrier density. Still, an abrupt change in the resistance could be observed when varying the SrO coverage from 0.83 to 1.0. The conductivity behavior changes dramatically from metallic-like to insulating and, therefore, no carrier density and mobility could be determined for a SrO coverage of precisely 100%. This suggests that a very well controlled growth of precisely 1 monolayer of SrO is necessary to fabricate the insulating AlO$_2$–SrO interface. At 0.83 of a monolayer of SrO the probability of electrons to find a percolation path to induce conductivity is apparently still too high.

In the case of a LaAlO$_3$ film on a TiO$_2$-terminated surface Ohtomo and Hwang claimed that a high-mobility electron gas was present at the LaO-TiO$_2$ interface.[4] They showed that the temperature dependence of the sheet resistance for a 60 Å thick LaAlO$_3$ layer on SrTiO$_3$ (LaO-TiO$_2$ interface) varies for different oxygen pressures during growth[^2], see **Fig. 8**, although thicker LaAlO$_3$ films showed very little oxygen pressure dependence. The temperature dependence of the Hall coefficient $R_H$ as reported in ref. 4 is given in Figure 8b

---

[^2]: The incorrect y-axis label in the original paper of Ohtomo and Hwang [4] was later corrected from 'mΩ' to 'Ω' in a corrigendum [61], see Figure 8a.

and there is little or no evidence for carrier freeze-out in most samples, although the interface grown at the highest oxygen pressure of $10^{-4}$ Torr exhibits a small increase in carrier density at higher temperatures. The resultant Hall mobility $\mu_H$ is given in Figure 8c demonstrating the extremely high carrier mobility that can be obtained at the interface for the samples grown at the lowest $pO_2$ value of $10^{-6}$ Torr. However, interfaces grown at higher oxygen pressures clearly display a much lower mobility.

The thickness dependence of the transport properties was investigated by Thiel *et al.* [62] who grew ultrathin $LaAlO_3$ layers of a few unit cells on a $TiO_2$-terminated $SrTiO_3$ substrate at an oxygen pressure of $2\times10^{-5}$ mbar. They found an abrupt transition from insulating to metallic behavior with a critical thickness of 4 unit cells above which the interfaces were conducting, see **Fig. 9**. In an earlier study Huijben *et al.* [63] found that single unit cell $LaAlO_3$ layers in $SrTiO_3/LaAlO_3/SrTiO_3$ heterostructures still showed metallic behavior. In such heterostructures two closely spaced complementary interfaces ($LaO-TiO_2$ and $AlO_2-SrO$) are present, which are electronically coupled. A critical separation distance of six perovskite unit cell layers, corresponding to approximately 23 Å, was found below which a decrease of the interface conductivity and carrier density occurs, see **Fig. 10**. Interestingly, the high carrier mobilities (~1000 $cm^2V^{-1}s^{-1}$ at low temperatures) characterizing the separate conducting interfaces were found to be maintained in coupled structures down to sub nanometer interface spacing.

Field effect devices based on $LaAlO_3-SrTiO_3$ heterostructures have been fabricated in order to investigate the interface transport properties as function of applied electric field.[62] Thiel *et al.* demonstrated memory behavior in the field effect structure, where they alternatingly applied a positive and negative gate voltage across the $SrTiO_3$ substrate to reversibly switch the sheet conductance three orders of magnitude, see **Fig. 11a**. Depending on the applied voltage (e.g.

~100 Volt over a period of time), these devices can be used as well to induce migration of oxygen in the SrTiO$_3$ substrate.

A related device concept was demonstrated by Cen *et al*.[64] by writing and erasing of nanowires in these structures. They showed the possibility to 'write' and 'erase' conducting wires between two electrodes with the tip of an atomic force microscope, which could be the first step towards highly dense nanodevices, see **Fig. 11b**. The observed conductive switching was ascribed by Cen *et al*. by local modulation of the oxygen stoichiometry in the topmost LaAlO$_3$ surface layer, which could be accompanied by accumulation of mobile electrons at the interface. However, possible other contributions such as charging of trap states and surface contamination have to be taken into account as well.[65]

**3.2. Superconductivity**

Superconductivity was observed in LaAlO$_3$-SrTiO$_3$ heterostructures by Reyren *et al.*.[66] They deposited LaAlO$_3$ layers with thicknesses of 8 and 15 unit cells (uc) on TiO$_2$-terminated SrTiO$_3$ substrates at an oxygen pressure of 6×10$^{-5}$ mbar. The films were additionally cooled to room temperature in 400 mbar of O$_2$ with a 1-hour oxidation step at 600°C. Subsequently, bridges were patterned with widths of 100 μm and lengths of 300 μm and 700 μm for four-point measurements.[67]

The 8-uc and 15-uc samples underwent a transition into a state for which no resistance could be measured at respectively ≅ 200 mK and ≅ 100 mK, see **Fig. 12a**. Application of a magnetic field μ$_0$H=180 mT perpendicular to the sample completely suppresses this zero-resistance state (Figure 12b and 12c). Voltage versus current (V-I) characteristics of a bridge in the 8-uc sample displayed a well-defined critical current I$_c$ at low temperatures, see Figure 12d and 12e. The transition into the superconducting state is interpreted in terms of a Berezinskii-Kosterlitz-Thouless (BKT) transition from the measured $V \propto I^a$ power-law dependence as well as specific $R(T)$ characteristics. Reyren *et al.* raised the question whether the bulk of the SrTiO$_3$ substrate was superconducting or only a thin sheet at the interface (~10 nm), because the observed superconducting transition temperatures fall in the same range as for oxygen-deficient SrTiO$_{3-x}$.[68-70] They concluded that the observation of both superconducting and insulating behavior on the same sample is very hard to reconcile with a pure oxygen vacancy scenario (see also section 4.2).

## 3.3. Magnetism

SrTiO$_3$-LaAlO$_3$ interfaces were found to exhibit magnetic effects as well.[71,72] This result is surprising since neither of the constituent compounds, SrTiO$_3$ and LaAlO$_3$, is magnetic. When LaAlO$_3$ is deposited under sufficiently high oxidation circumstances (i.e. above $10^{-3}$ mbar) two conclusions were drawn based on the transport properties of the interfaces,[71] see **Fig. 13a**. First of all, localized magnetic moments are present. Secondly, a coupling exists between the magnetic moments at very low temperatures (order of 300 mK).

The presence of localized magnetic moments is evidenced from a number of observations.[71] At low temperatures, a large negative magnetoresistance is observed that is independent of field orientation, see Figure 13b. Additionally, the resistance increases logarithmically with decreasing temperature below a temperature of around 70 K. A logarithmic increase in resistance is known to occur for 2D weak localization as well. In the case of 2D weak localization, however, the magnetoresistance is a function of the enclosed flux and should thus depend on the orientation of the field. The orientation dependence is absent for 3D weak localization, but in this case no logarithmic dependence is expected. Therefore, the negative magnetoresistance, as well as the logarithmic temperature dependence, are suggested to arise from scattering at localized magnetic moments. Many different scattering models can explain the negative magnetoresistance, while especially the Kondo model is suitable for explaining the logarithmic temperature dependence.[73]

In the presence of localized magnetic moments, the interesting question arises whether or not the moments can couple (anti)-ferromagnetically. A direct magnetization measurement is very challenging because of the presence of only a few moments at the interface. However, more indirectly, two indications for magnetic ordering are reported.[71] The magnetization of the interfaces can indirectly be deduced from the magnetoresistance. The derived susceptibility is found to follow a Curie-Weiss behavior, suggestive for the existence of a coupling between the moments at very low temperatures. In addition, below 300 mK hysteresis in the magnetoresistance is observed. Although the hysteresis is not directly evidence for ferromagnetism, it is indicative for a delayed response to an external signal presumably due to coupling between the magnetic momenta. The size of the hysteresis loop mainly depends on the magnetic field sweep rate. This sweep rate dependence is explained by the observed very long relaxation time (order of seconds at 300 mK).

## 4. Transport mechanisms

### 4.1. Structural aspects

Despite the possibility of valence changes, cation disorder can be present at oxide interfaces.[5] Additionally, structural deformations at $SrTiO_3$-$LaAlO_3$ interfaces can be expected from the strain. Pseudomorphically growing a thin film of $LaAlO_3$ on $SrTiO_3$ produces a $LaAlO_3$ layer which is in-plane tensile strained to the $SrTiO_3$ substrate and therefore shortened in the out-of-plane direction, see Section 2.3. However, to understand and model the properties of the interfaces, the positions of the atoms at the interface have to be determined accurately. Ab-initio structure calculations of this type of interfaces have already indicated that considerable atomic displacements will occur.[8-17]

The structure of the LaAlO$_3$/SrTiO$_3$ interface has been studied by transmission electron microscopy (TEM) [5,63,66,74-78] and surface X-ray diffraction.[79,80] Thin films of LaAlO$_3$ have been pseudomorphically grown on SrTiO$_3$ substrates up to thicknesses of ~20 nm. Although the LaAlO$_3$ film can be coherently grown on the SrTiO$_3$ substrate without any defects, still some TEM studies have demonstrated the existence of dislocations/point defects at the interface. To explain the different observations, one has to focus on the oxygen pressures and the layer thicknesses that were used during the LaAlO$_3$ growth in all studies. As mentioned in section 2.3 a transition from 2-dimensional layer-by-layer growth to island growth is present for oxygen deposition pressures above 10$^{-3}$ mbar. Maurice *et al.* [77] also showed that a 5 nm LaAlO$_3$ layer grown at 40 Pa (or 4×10$^{-1}$ mbar) had a very rough surface, a large number of dislocations and a fully relaxed crystal structure. They indicated that this high-pressure sample was insulating. On the other hand, Kalabukhov *et al.* [76] have demonstrated that LaAlO$_3$ thin films grown at very low oxygen pressures of 10$^{-6}$ mbar also contain a large number of dislocations at the interface, which are approximately ~15 nm apart. At these very low pressures the kinetic energy of the arriving species during pulsed laser deposition creates irradiation damage at the substrate surface and forms dislocations/point defects in the growing layer. In case of 2-dimensional layer-by-layer LaAlO$_3$ growth in the oxygen pressure range between 10$^{-5}$ and 10$^{-3}$ mbar the LaAlO$_3$/SrTiO$_3$ interface will be defect-free, as demonstrated by various groups.[5,63,66]

Detailed analysis of both types of possible atomic stackings (LaO-TiO$_2$ and AlO$_2$-SrO) at the LaAlO$_3$/SrTiO$_3$ interface by Nakagawa *et al.* [5], using angular dark field (ADF) imaging in scanning transmission electron microscopy (STEM), revealed that a small amount of atomic interdiffusion was present at the interface in both cases, but most significantly for a LaO–TiO$_2$ stacking. However, this atomic interdiffusion is very small and both types of interfaces can still clearly be distinguished in a LaAlO$_3$/SrTiO$_3$ heterostructure as shown by Huijben *et al.*.[63]

Although analysis of the total LaAlO$_3$ layer indicated a shortened c-axis (~3.73 Å) due to the tensile strain, atomic positions close to the interface could be very different due to the polar discontinuity. Maurice *et al.* used aberration-corrected high-resolution transmission electron microscopy (HRTEM) to study the local atomic structure and measured a dilation of the (00l) inter-planar distance at the interface.[74,75] They indicated that the unit cells at the interface are elongated by 4 - 9% (3.94 - 4.13 Å) from the bulk value and situated between the SrO and LaO planes at the interface, i.e. on the TiO$_6$ interfacial octahedral. Their explanation for this intrinsic distortion of the unit cells at the interface was a lowering of the electron energy by introducing an electron in the empty Ti-3d levels by an effect similar to Jahn-Teller.

Surface x-ray diffraction (SXRD) is a well-established technique for high-resolution structure determination of surfaces and interfaces. To study the atomic heterointerface structure during its formation by pulsed laser deposition, Vonk *et al.*[79] performed an in-situ SXRD study at the deposition conditions. They observed no clear atomic displacements from the ideal bulk STO lattice sites at the deposition temperature of 1123 K. However, at a lower temperature of 473 K the distortions become significant, whereby the anions displace towards and the cations away from the underlying SrTiO$_3$ substrate. The interatomic distances across the interface between the cations are in the range 4.0 – 4.1 Å, which is very comparable to results by HRTEM studies.[74,75] The opposite displacements of cations and anions, seen as strong buckling of the atomic layers, result in the TiO$_6$ octahedra at the interface to contract their axis in the surface normal direction. These results could originate from a Jahn-Teller effect: the initially unoccupied d-shells of one-half of the interface Ti atoms, receive one electron. However, it should be realized that large electronic reconstruction effects are only to be expected when more than a single unit-cell layer is deposited. In an ex-situ SXRD study Wilmott *et al.* [80] also observed a dilation at the interface for a layer of 5 unit cells. They also indicated that cation intermixing was observed at a greater depth for Sr/La than Ti/Al.

**4.2. Oxygen vacancies**

*4.2.1. Low pressure samples*

In the previous section, it was already mentioned that pulsed laser deposition of LaAlO$_3$ at very low oxygen pressures of $10^{-6}$ mbar will lead to defects at the LaAlO$_3$/SrTiO$_3$ interface due to irradiation damage at the SrTiO$_3$ substrate surface. The high kinetic energy of the arriving species will create oxygen vacancies in the top layer of the SrTiO$_3$ substrate, which is subsequently protected when a full LaAlO$_3$ layer has been grown. Oxygen vacancies can easily be formed in SrTiO$_3$ [81], due to the change in valence of the Ti-atom from Ti$^{4+}$ to Ti$^{3+}$. Samples grown at such low oxygen pressures of $10^{-6}$ mbar change color from transparent to grey/black, which is characteristic for oxygen reduced SrTiO$_3$.[82]

Ohtomo and Hwang already indicated that for LaAlO$_3$/SrTiO$_3$ interfaces grown at the lowest oxygen pressures of $10^{-6}$ Torr an interpretation of *1/R$_H$e* as a sheet carrier density would imply 'unphysical densities' ($10^{17}$ cm$^{-2}$ requires a unit cell carrier density of $1.7 \times 10^{22}$ cm$^{-3}$ over 600 Å of thickness). At the same time they observed unusually high mobilities of ~$10^4$ cm$^2$V$^{-1}$s$^{-1}$. These two observations can only be explained by including the possible role of oxygen vacancies in the top layer of the SrTiO$_3$ substrate, despite annealing in molecular oxygen.

The effect of oxygen vacancies in the SrTiO$_3$ substrate on the electrical properties of the LaAlO$_3$/SrTiO$_3$ interface were investigated in more detail by Kalabukhov *et al.* [76], Basletic *et al.* [83] and Siemons *et al.*. [84,85] Kalabukhov *et al.* [76] demonstrated by cathode luminescence and photoluminescence that LaAlO$_3$-SrTiO$_3$ heterostructures grown at low oxygen pressure ($10^{-6}$ mbar) displayed the same light intensity and wavelength, as intentionally oxygen-reduced $SrTiO_{3-x}$ by Ar-ion bombardment or vacuum annealing. Luminescent light was also observed from samples grown at $10^{-4}$ mbar, though with much weaker intensity. Basletic *et al.*[83] performed resistance profile mapping in cross-section samples with a conducting-tip atomic force microscope. They confirmed the occurrence of conductivity through the bulk of $SrTiO_3$ in samples grown at low oxygen pressures ($10^{-6}$ mbar), which is in agreement with their earlier magnetotransport experiments.[86] For samples grown at the same low oxygen pressure, but subsequently annealed at 300 mbar to minimize oxygen vacancies, evidence was given for the presence of a conductive region confined within ~7 nm next to the interface. Siemons *et al.* [84,85] performed ultraviolet photoemission spectroscopy (UPS), near edge x-ray absorption spectroscopy (NEXAS) and visible to vacuum ultraviolet spectroscopic ellipsometry (vis-VUV-SE) measurements on samples prepared under various oxidation conditions. They showed a strong dependence of the properties of the conducting layer on the oxidation conditions during growth, which they correlated, through the bonding conditions of the Ti-atom, to oxygen vacancies in $SrTiO_3$. Further study was done by annealing in atomic oxygen, while the possibility of introducing interstitial oxygen was minimized. Although the carrier density could be drastically reduced, a lower limit value independent of temperature was observed for anneals above 500 °C, while the mobility stayed constant; a summary of these experiments can be found in **Fig. 14**. Finally, they suggested that the observed high mobilities in the presence of a large amount of defects/oxygen vacancies can be explained by a simple model in which the charge carriers move away from the defect layer, where they originated, into the pristine $SrTiO_3$ crystal and therefore experience less scattering.

*4.2.2. Superconductivity in $SrTiO_3$*

SrTiO$_3$ can be electron doped by replacing only a small fraction of Sr with Nb, La or Ta and will become highly conducting with a carrier density of about $10^{19}$-$10^{20}$ cm$^{-3}$ and even superconducting below 400 mK.[68-70,88] It is suggested that the role of the Nb doping is to create oxygen vacancies that donate charge to the lattice, similar to intentionally oxygen-reduced SrTiO$_{3-x}$. In this context, it is worthwhile to revisit the observed superconductivity at the SrTiO$_3$-LaAlO$_3$ interface.[66] Reyren *et al*. rightly stated that if the superconductivity were due to oxygen defects in SrTiO$_{3-x}$, a carrier density of $\gtrsim 3\times10^{19}$ cm$^{-3}$ would be required for a $T_c$ of 200 mK. They measured sheet carrier densities of about 1.5-4×10$^{13}$ cm$^{-2}$, which would give an upper limit for the thickness of the superconducting sheet of $\cong$ 15 nm. If calculations of the carrier density profile at interfaces in oxygen-deficient SrTiO$_{3-x}$ [84] are used a sheet carrier density > 5×10$^{14}$ cm$^{-2}$ is needed to provide a carrier concentration of 3×10$^{19}$ cm$^{-3}$. Therefore, they concluded that the superconductivity of the LaAlO$_3$/SrTiO$_3$ interface cannot be caused by doped SrTiO$_{3-x}$ doping alone.

*4.2.3. Magnetic effects from oxygen vacancies*

For LaAlO$_3$ deposited on SrTiO$_3$ at relatively high pressures (10$^{-3}$ mbar) localized magnetic moments have been observed [71], giving rise to a large negative magnetoresistance and a logarithmic upturn in the temperature dependence of the interface sheet resistance. Below 300 mK magnetic coupling has been observed with a characteristic long time scale. It is known, that alloying SrTiO$_3$ with magnetic ions such as Co and Cr can give magnetic effects [89,90] when the dopant concentration is sufficiently high (order of 10%), while doping with non-magnetic ions has not yet provided a means of inducing magnetism in SrTiO$_3$. X-ray photoelectron spectroscopy on the magnetic interfaces has ruled out measurable amounts of magnetic dopants,[71] but cation vacancies and oxygen vacancies cannot be excluded out beforehand.[91]

In general, in order to induce magnetism from non-magnetic ions, cation vacancies, or oxygen vacancies, one needs to estimate the thermodynamic stability of such a dopant, combined with the exchange energy and necessary percolation threshold. It has, for example, been predicted that a Ca vacancy in CaO could provide magnetism, while the oxygen vacancy was found not to provide magnetic states.[92] More recently, a substitution of nitrogen for oxygen has been suggested to be able to provide magnetism in otherwise non-magnetic oxides [93] and first indications exist for cation vacancy induced magnetism in Nb doped $TiO_2$.[94] From the absence of any other known material in which oxygen vacancies give rise to magnetism, and the trend that magnetic effects at the $SrTiO_3/LaAlO_3$ interface become strong when the amount of oxygen vacancies is diminished, we conclude that the magnetic effects can likely be attributed to the electronic interface reconstruction instead.

### 4.3. Electronic interface reconstruction

From the discussion above it becomes apparent that structural effects, such as dilations and cation disorder, as well as oxygen vacancies can give rise to (super)conducting behavior at $SrTiO_3$-$LaAlO_3$ interfaces. However, it has also been shown how interfaces with relatively low cation disorder can be grown and how the influence of oxygen vacancies can be diminished [65,71] by raising the oxygen background deposition pressure during growth. The emerging picture shows a transition from samples completely dominated by oxygen defects, grown at $10^{-6}$ mbar, to samples in an intermediate regime that become superconducting, to a regime where localized magnetic moments are observed (grown at $10^{-3}$ mbar), see **Fig. 15**.

We suggest that electronic reconstruction is a significant factor in the transport properties for interfaces grown at high enough oxygen partial pressure. There are some experimental observations that support this idea and that are difficult to reconcile with pure oxygen defect scenarios. First of all, a strong influence of the substrate termination on the conductivity is observed. This is shown in figures 6 and 7. Secondly, the modulation of carrier density by coupling adjacent interfaces [63] and the sharp conducting onset on the crossover from 3 to 4 deposited unit cells of LaAlO$_3$ [62] is more difficult to attribute to the oxygen vacancy scenario than electronic interface reconstruction. Theoretical support for this argument is provided by ab-initio calculations that show how the carrier density evolves with layer thickness. Finally, it was argued in section 4.2 that the localized magnetic moments are likely to originate from other mechanisms rather than oxygen vacancies.

## 5. Conclusion and outlook

The emerging picture that describes the relation between structure and property of the SrTiO$_3$-LaAlO$_3$ interface is one in which the variety of observed transport properties can be explained by the relative contribution of three structural aspects: oxygen vacancies, structural deformations (including cation disorder) and electronic interface reconstruction. The resulting temperature vs. doping phase diagram is shown in **Fig. 16**. The doping scale consists of the intrinsic carrier doping from the electronic reconstruction ($x_{int}$), carrier doping from oxygen vacancies ($x_{O2}$) and carrier doping by applied electric fields ($x_{field}$). The exact positions of $x_{int}$ and the 2D-3D crossover in the phase diagram are not yet known and can vary due to several other polarity discontinuity compensating mechanisms (such as $x_{O2}$). When the phase diagram of Figure 16 is compared to related bulk oxide phase diagrams, such as for Nb doped SrTiO$_3$[88], it becomes apparent that the richness of the first is due to the additional interface reconstruction.

The SrTiO$_3$-LaAlO$_3$ interface has only been an example in our review to demonstrate the richness of oxide interface transport properties (and the complexity of the structure-property relationship). Naturally, progress has been made towards other oxide interface systems, using a similar simplified picture of polar discontinuities to identify these systems. A recent example is the conducting $SrTiO_3/LaVO_3$ interface [95,96], where the amount of carriers is close to the expected value from polar effects. Of course, in the $SrTiO_3/LaVO_3$ case, one has to consider two possible multivalent ions, being the Ti and the V. Another development, where polar effects could play an important role is the field of high-temperature superconducting (HTS) cuprates, as pointed out by Koster *et al.* [97], possibly providing new insight in the observations of unexpected superconductivity occurrence [98] and $T_c$ enhancements.[99,100]

Concluding, interface reconstruction is an important phenomenon that can occur in many complex oxide heterostructures, providing an extension of the exceptional oxide electronic properties. Irrespective of the origin of mobile charge at these heterointerfaces, once harnessed and controlled all could lead to interesting heterostructures, where the properties are determined by strong correlation effects, and therefore quite unpredictable and full of surprises.

The authors acknowledge fruitful discussions and interactions with M.R. Beasley, J. Chakhalian, T. Claeson, T.H. Geballe, H.Y. Hwang, J.C. Maan, J. Mannhart, R. Pentcheva, W.E. Pickett, J.-M. Triscone, S. Van Tendeloo, T. Venkatesan & U. Zeitler and their respective group members. Furthermore our group members J. Huijben, W. Siemons, W. Van der Wiel & M. Van Zalk are acknowledged for their scientific contributions.

This work is part of the research program of the Foundation for Fundamental Research on Matter (FOM, financially supported by the Netherlands Organization for Scientific Research (NWO)), by the EU through Nanoxide, by the European Science Foundation through THIOX, and by NanoNed, Dutch initiative on nanotechnology.

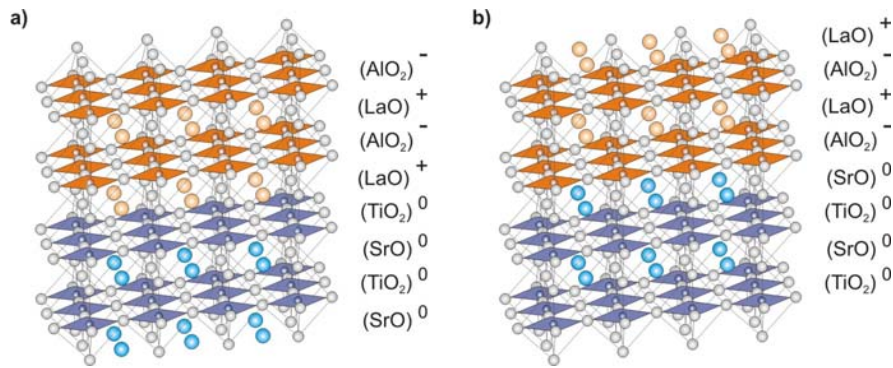

**Figure 1**. Schematic models of the two possible interfaces between SrTiO$_3$ and LaAlO$_3$ in the (001)-direction. The resulting (LaO)$^+$/(TiO2)$^0$ (a) and (AlO2)$^-$/(SrO)$^0$ interfaces (b), showing the composition and the ionic charge state of each layer. The schematic models are taken from ref. 4.

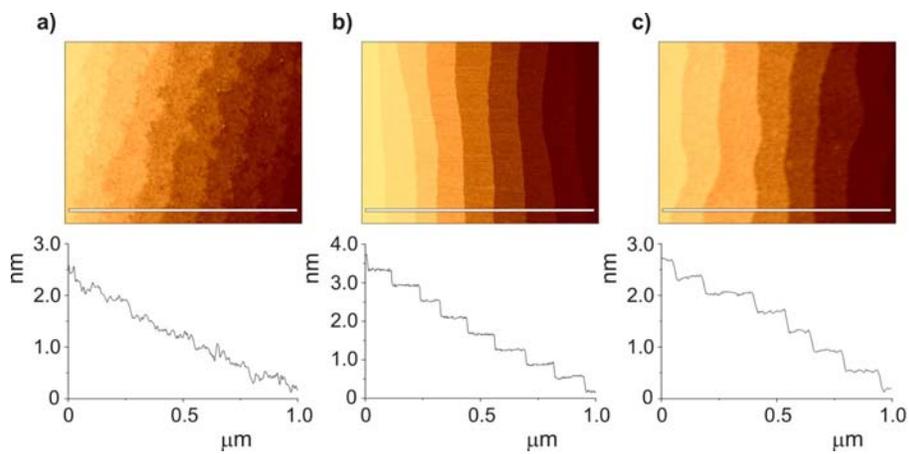

**Figure 2**. Surface analysis of SrTiO$_3$ substrates by atomic force microscopy. AFM micrograph and surface roughness analysis result of an as-received (ethanol cleaned) double-terminated surface (a), a chemically and thermally treated single TiO$_2$-terminated surface (b) and a pulsed laser deposited single SrO-terminated surface (c).

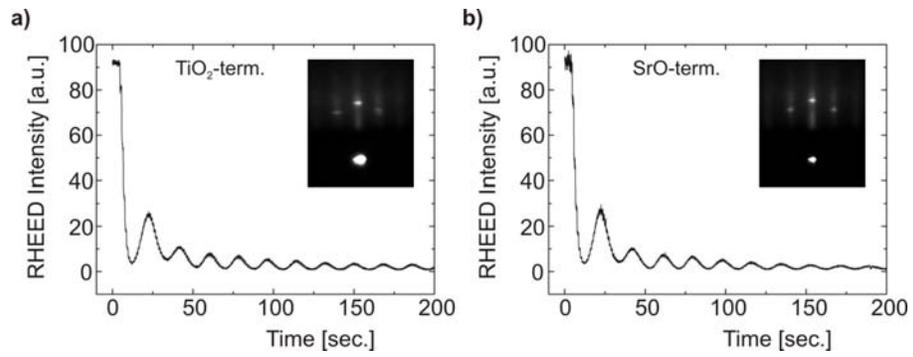

**Figure 3**. Monitoring of the RHEED intensity during initial growth of LaAlO$_3$ unit cells on single-terminated SrTiO$_3$ substrates with a TiO$_2$-terminated surface (a) and a SrO-terminated surface (b). In the insets are the RHEED patterns shown with the clear 2-dimenisonal RHEED spots after growth of 26 unit cells of LaAlO$_3$.

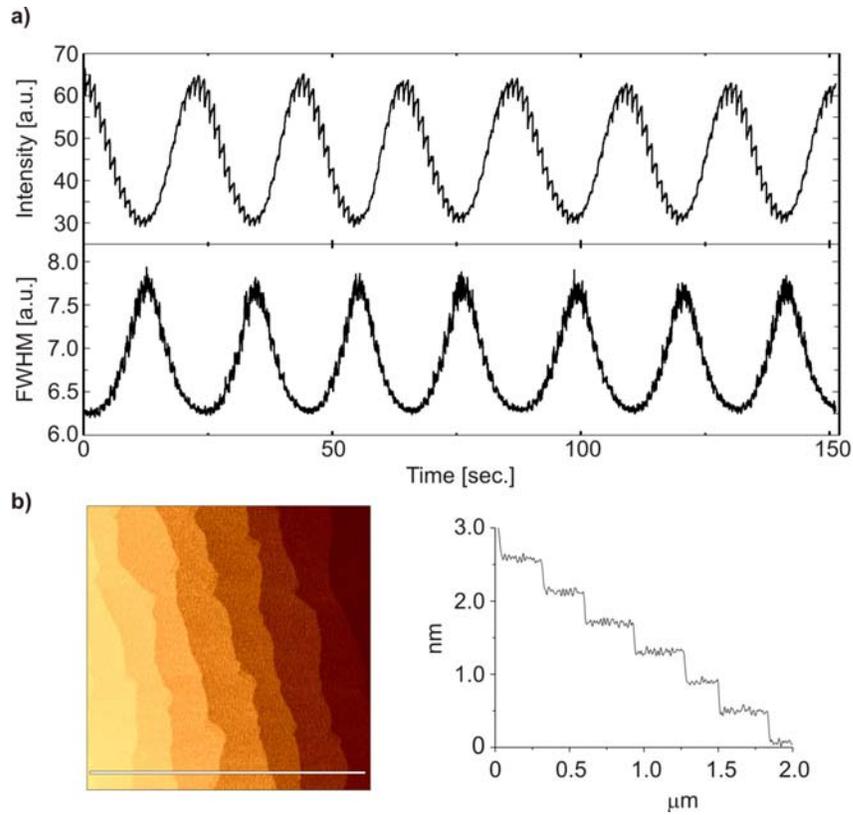

**Figure 4**. RHEED intensity and FWHM monitoring during growth of LaAlO$_3$ unit cells on a SrTiO$_3$ substrate (a) and surface analysis by atomic force microscopy of a 26 unit cells thick LaAlO$_3$ thin film on a TiO$_2$-terminated SrTiO$_3$ substrate (b). The roughness analysis shows smooth terraces with unit cell steps.

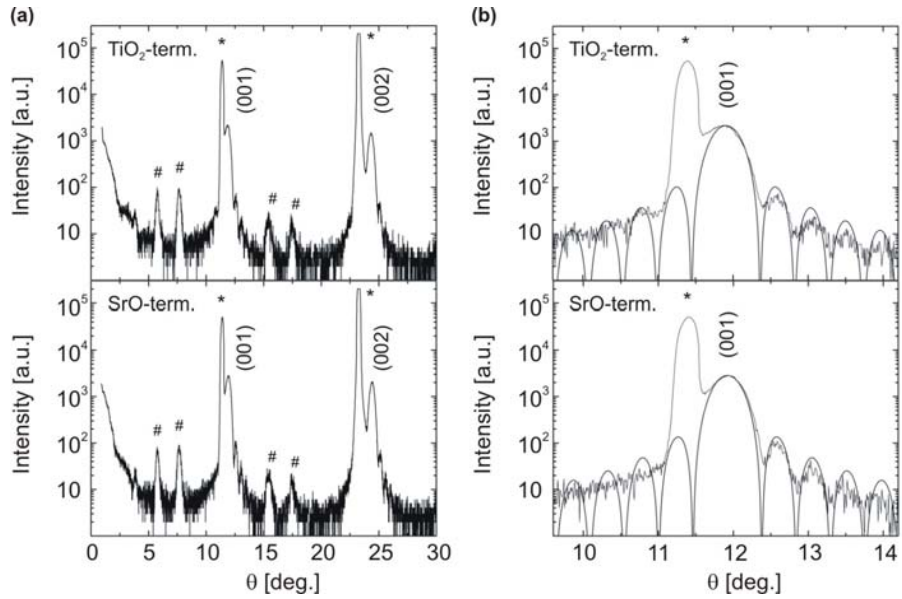

**Figure 5**. X-ray diffraction analysis of a 26 unit cells thick LaAlO$_3$ thin film on SrTiO$_3$ substrates with a TiO$_2$-terminated surface (top) and a SrO-terminated surface (bottom). Shown are large angle θ-2θ scans (a) as well as more detailed θ-2θ scans around the (001) reflections of the LaAlO$_3$ thin films (b). A fit to the Kiessig fringes in the detailed θ-2θ scans is also shown. The SrTiO$_3$ substrate reflections are indicated with an asterisk and their spectral contributions (λ/2 and λ/3) with a cross.

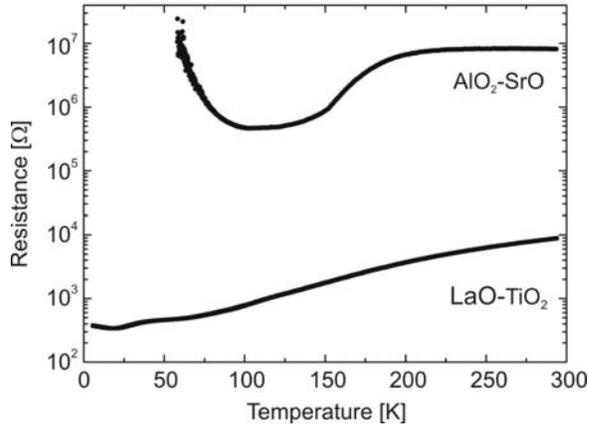

**Figure 6**. Temperature dependence of the resistance for 26 unit cells thick LaAlO$_3$ films on SrTiO$_3$ substrates with a TiO$_2$-terminated surface and a SrO-terminated surface both grown at 850 °C and 3×10$^{-5}$ mbar oxygen pressure.

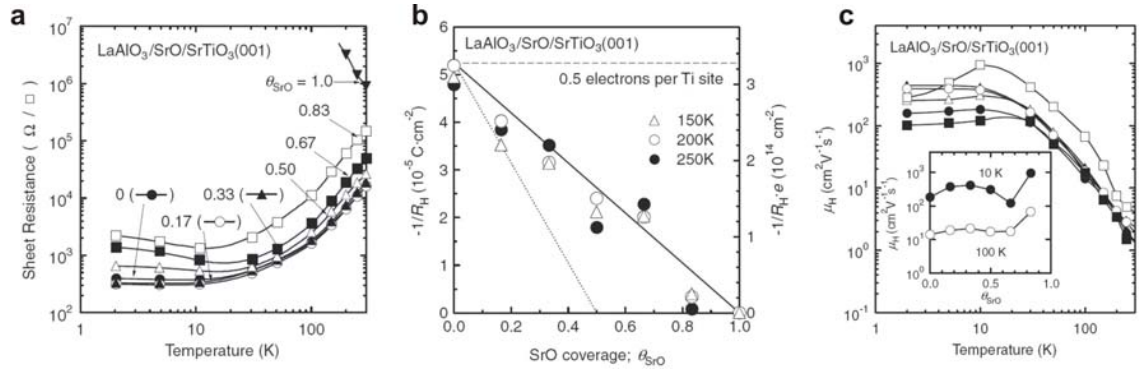

**Figure 7**. SrO fractional coverage ($0 \leq \theta_{SrO} \leq 1$) dependence of sheet resistance (a), inverse Hall coefficient $-1/R_H$ (b) and Hall mobility $\mu_H$ (c) for LaAlO$_3$/SrO/SrTiO$_3$ heterointerfaces. Figures are taken from ref. 60.

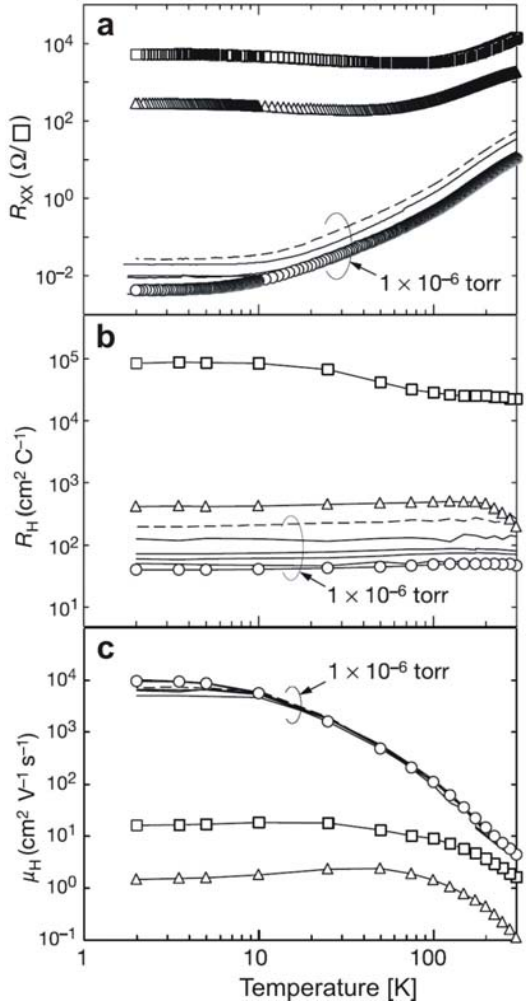

**Figure 8**. Transport properties of the $(LaO)^+/(TiO_2)^0$ interface for different oxygen partial pressures pO$_2$ during growth at $10^{-4}$ (□), $10^{-5}$ (Δ), and $10^{-6}$ (○) torr, as well as for $10^{-6}$ torr growth followed by annealing in 1 atm. of O$_2$ at 400 °C for 2 hours (dashed line). Temperature dependence of sheet resistance R$_{XX}$ (a), Hall Coefficient R$_H$ (b) and Hall mobility µ$_H$ (c) for the interface between 60 Å thick LaAlO$_3$ and SrTiO$_3$, respectively. Figures are taken from ref. 4.

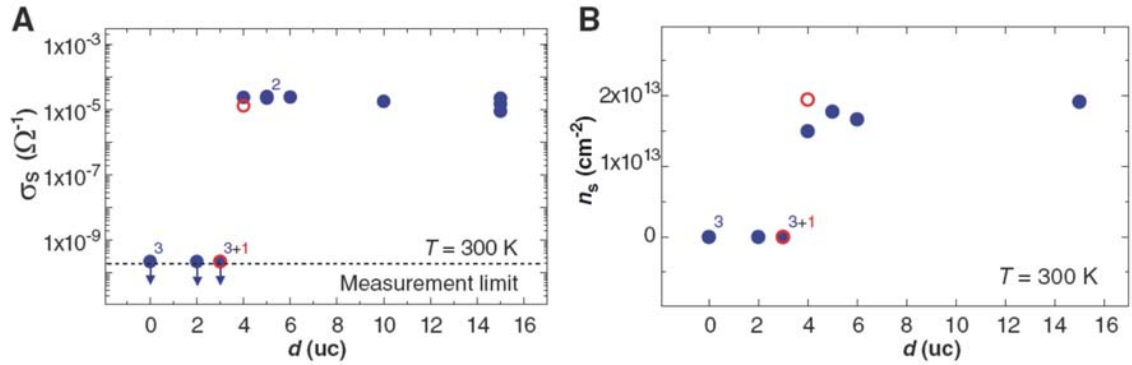

**Figure 9**. Influence of LaAlO$_3$ thickness on the electronic properties of the LaAlO$_3$/SrTiO$_3$ interfaces. (a) Sheet conductance and (b) carrier density of the heterostructures plotted as a function of the number of their LaAlO$_3$ unit cells. The data shown in blue and red are those of samples grown at 770°C and 815°C, respectively. The data were taken at 300 K. The numbers next to the data points indicate the number of samples with values that are indistinguishable in this plot. Figures are taken from ref. 62.

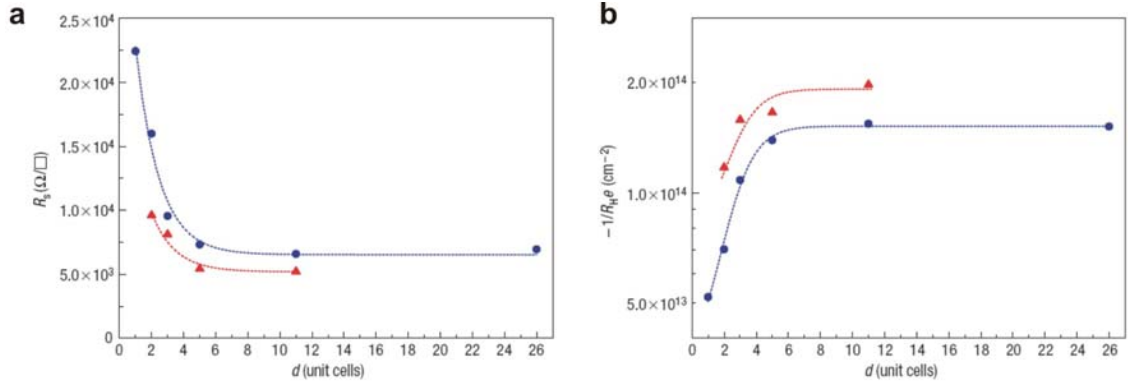

**Figure 10**. Electronically coupled complementary interfaces in LaAlO$_3$/SrTiO$_3$ heterostructures. (a) Dependence of the sheet resistance $R_S$ on the separation distance $d$. (b) Dependence of $-1/R_He$ on the separation distance $d$. SrTiO$_3$/LaAlO$_3$/SrTiO$_3$ heterostructures and LaAlO$_3$/SrTiO$_3$/LaAlO$_3$ heterostructures are indicated by circles and triangles, respectively. The dashed lines are guides to the eye. Figure is taken from ref. 63.

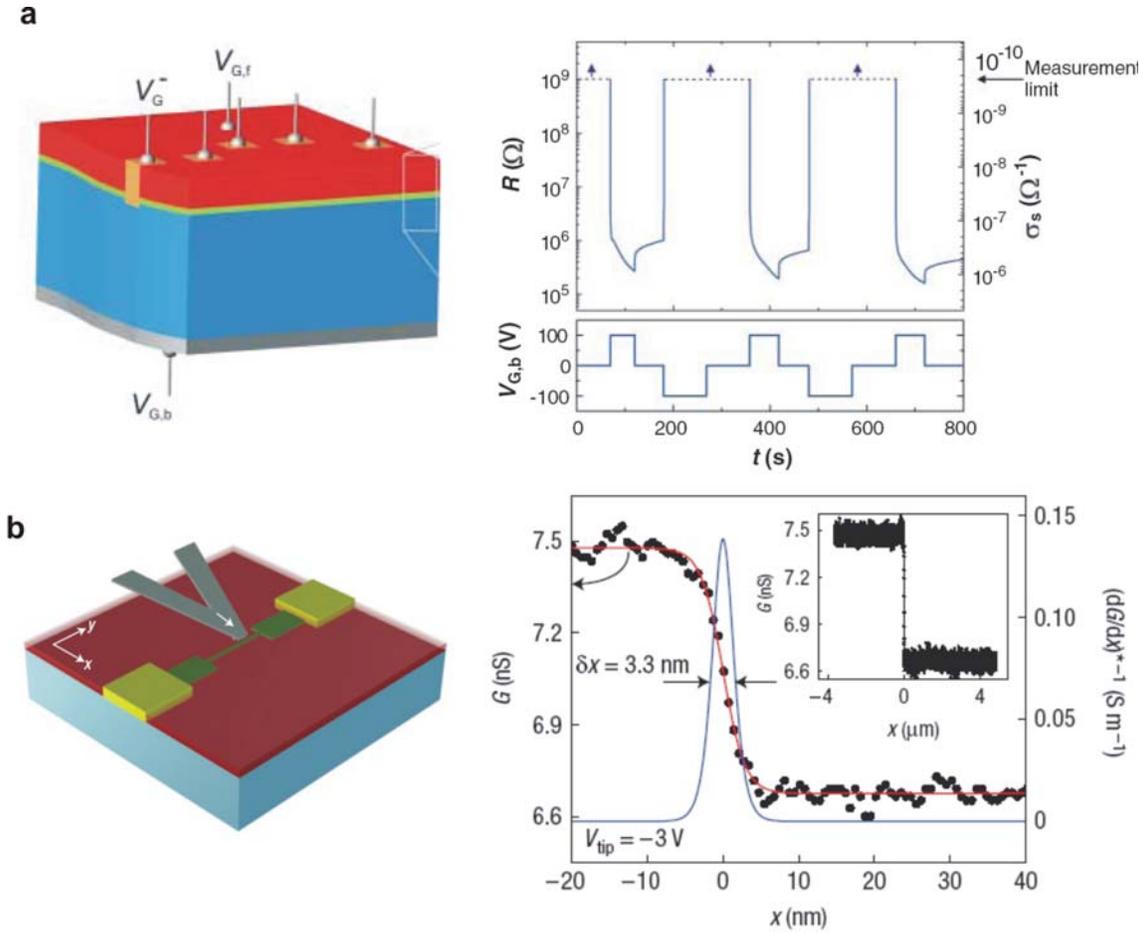

**Figure 11**. Device structures based on LaAlO$_3$-SrTiO$_3$ heterostructures. (a) Memory behavior: Sheet resistance measured at 300 K and applied backgate voltage, both plotted as a function of time for a LaAlO$_3$ layer with a thickness of 3 unit cells. By applying the gate voltage pulses, the sheet conductance could reversibly be switched between ~1×10$^{-6}$ ohm$^{-1}$ and an unmeasurable value <2×10$^{-10}$ ohm$^{-1}$. The data were measured in fourpoint configurations. Figure is taken from ref. 62. (b) Writing and erasing nanowires: Conductance between the two electrodes measured as a function of the tip position across the wire, while cutting the wire with the tip biased at −3 V. A sharp drop in conductance occurs when the tip passes the wire. The inset shows the conductance measured over the entire 8 μm scan length. Figure is taken from ref. 64.

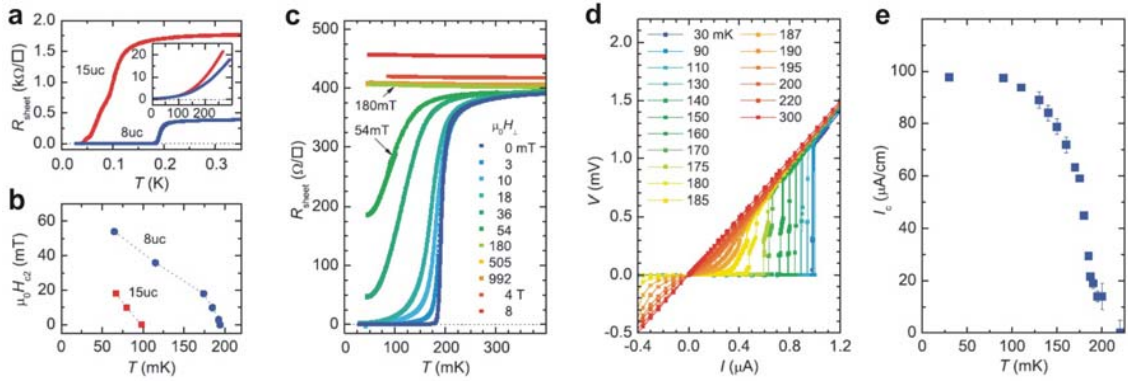

**Figure 12**. Transport measurements on LaAlO$_3$/SrTiO$_3$ heterostructures. (a) Dependence of the sheet resistance on T of the 8-uc and 15-uc samples (measured with a 100-nA bias current). (Inset) Sheet resistance versus temperature measured between 4 K and 300 K. (b) Temperature dependence of the upper critical field Hc$_2$ of the two samples. (c) Sheet resistance of the 8-uc sample plotted as a function of T for magnetic fields applied perpendicular to the interface. (d) Temperature-dependent voltage-current characteristics of a 100×300 mm$^2$ bridge of the 8-uc LaAlO$_3$/SrTiO$_3$ heterostructure. (e) Measured temperature dependence of the linear critical current density, as obtained from (d). Figures are taken from ref. 66.

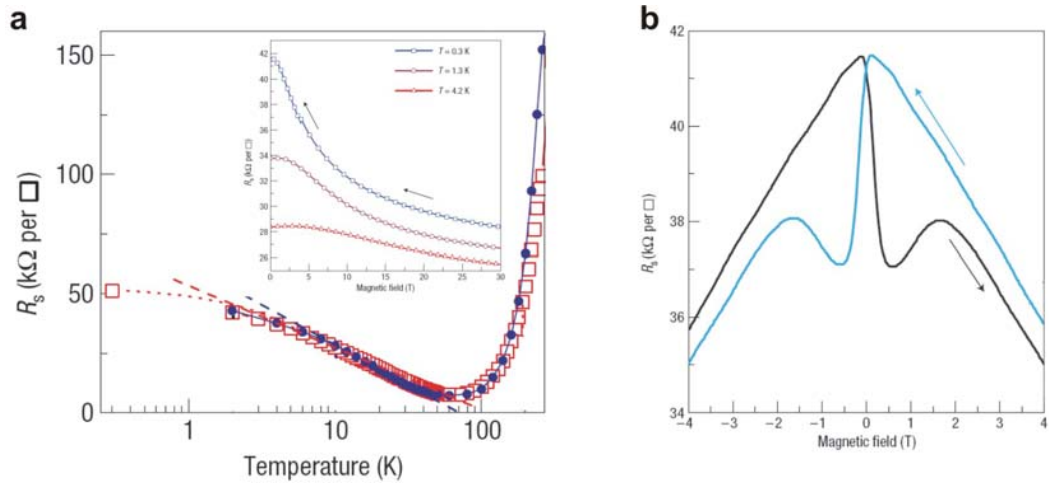

**Figure 13**. Magnetic ordering at LaAlO$_3$/SrTiO$_3$ interfaces. (a) Temperature dependence of the sheet resistance R$_S$ for two LaAlO$_3$/ SrTiO$_3$ conducting interfaces, grown respectively at a partial oxygen pressure of 2.5×10$^{-3}$ mbar (open squares) and 1.0×10$^{-3}$ mbar (filled circles). The low-temperature logarithmic dependencies are indicated by dashed lines. Inset: Large negative magnetoresistance in sheet resistance under applied magnetic field perpendicular to the interface at 0.3, 1.3 and 4.2 K. The magnetic-field sweep direction is indicated by arrows. (b) Sheet resistance at 0.3 K of a SrTiO$_3$/LaAlO$_3$ conducting interface, grown at 1.0×10$^{-3}$ mbar. The arrows indicate the direction of the measurements (at a rate of 30mTs$^{-1}$). Figures are taken from ref. 71.

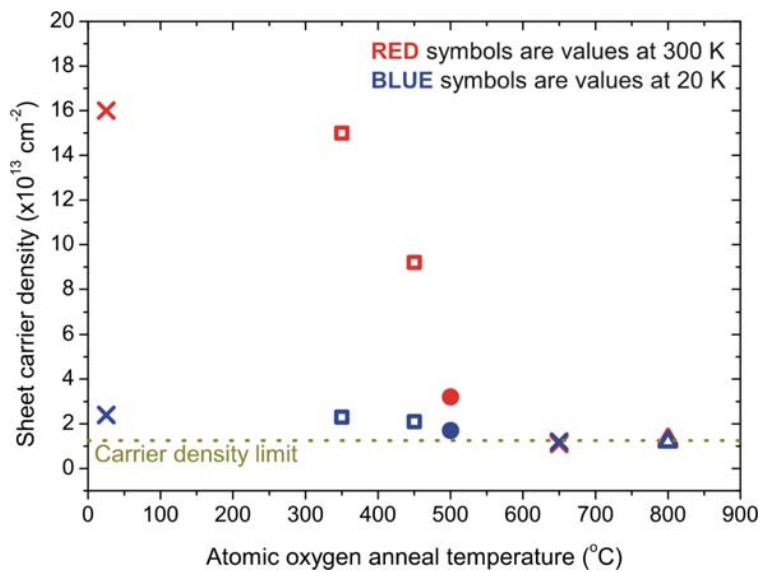

**Figure 14**. Sheet carrier densities at 20 K (blue symbols) and 300 K (red symbols) as a function of annealing temperature in 600 W atomic oxygen, for samples made at $10^{-5}$ Torr of $O_2$ (600 W of atomic oxygen corresponds to ~$10^{17}$ oxygen atoms cm$^{-2}$ s$^{-1}$). The latter value was taken after Ingle *et al*. [87] who worked on the same system in the same laboratory. The values at 25 °C indicate the as deposited samples. The different symbol shapes indicate different samples made under similar conditions: two made at Stanford (circles and triangles), the other two made at the University of Twente (crosses and squares). Sample thickness ranges from 5 to 26 ML. Figure is taken from ref. 84.

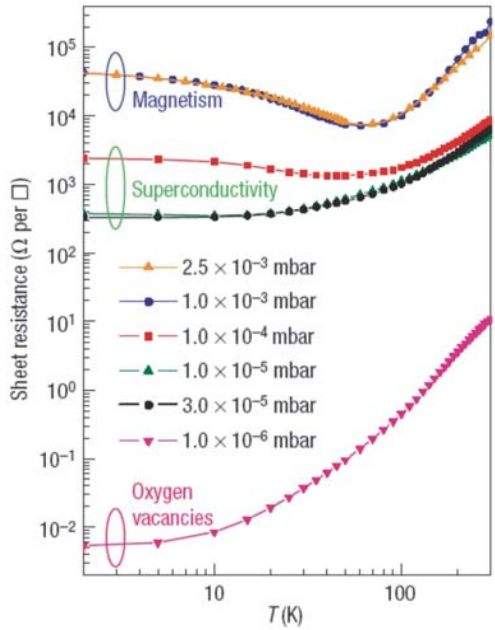

**Figure 15**. Sheet resistance of n-type $SrTiO_3/LaAlO_3$ interfaces. Temperature dependence of the sheet resistance for $SrTiO_3/LaAlO_3$ conducting interfaces, grown at various partial oxygen pressures (data from ref. 71). Three regimes can be distinguished: low pressures leads to oxygen vacancies, samples grown at high pressures show magnetism, whereas samples grown in the intermediate regime show superconductivity. Figure is taken from ref. 65.

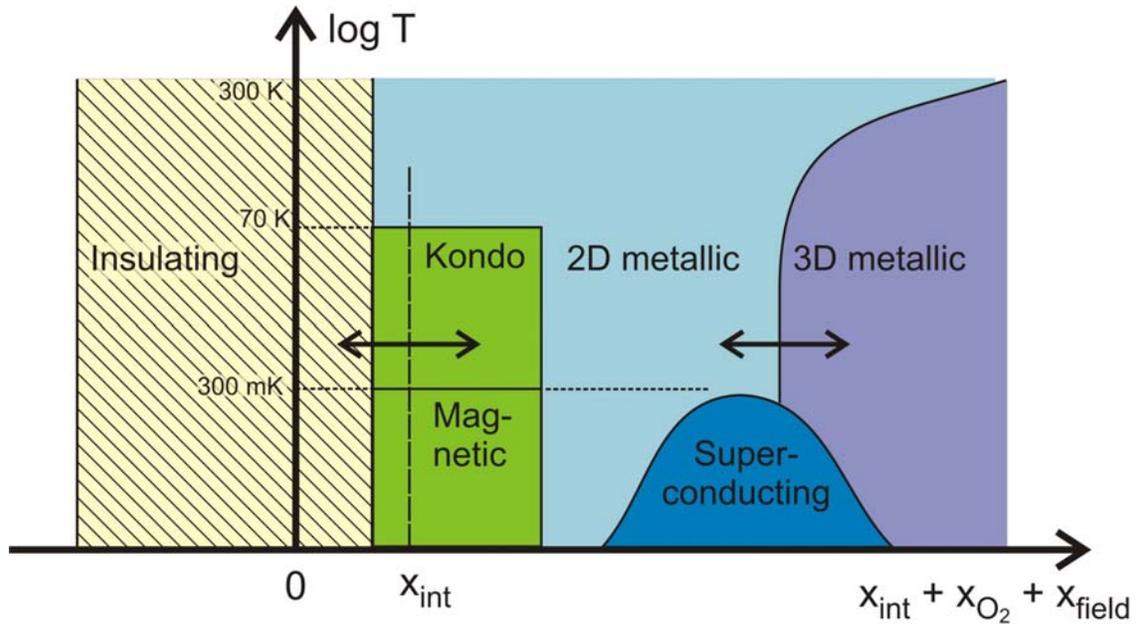

**Figure 16**. Doping vs. temperature phase diagram of $SrTiO_3$-$LaAlO_3$ interfaces. The doping scale consists of three possible contributions: intrinsic carrier doping from the electronic reconstruction ($x_{int}$), carrier doping from oxygen vacancies ($x_{O2}$) and carrier doping by applied electric fields ($x_{field}$). Observed transport effects are insulation at p-type interfaces (x<0) [4], 2D interface metallicity [4], 3D bulk metallicity [4,83], Kondo effect around $T_K$ = 70 K [71], magnetism below 300 mK [71], and superconductivity below $T_c$ = 200 mK [66]. The exact position of $x_{int}$ in the phase diagram is not yet known and can vary due to other polarity discontinuity compensating mechanisms (such as $x_{O2}$). The 2D-3D transition in the metallicity can be calculated from electrostatics (as in ref. 84) and lies generally at lower carrier densities for lower temperatures.